\documentclass[aps,prl,twocolumn,showpacs]{revtex4}
\usepackage{amsmath}
\usepackage{amssymb}
\usepackage{graphics}

\newcommand\Ho{\hat H}
\newcommand\bo{\hat b}
\newcommand\xist{\xi^\ast}
\newcommand\dg{^\dagger}
\newcommand\pr{^\prime}
\newcommand\astar{a^\ast}
\newcommand\om{\omega}
\newcommand\so{\hat s}
\newcommand\const{\rm{const}}
\newcommand\e{\mathrm{e}}
\newcommand\hc{\mathrm{h.c.}}

\newcommand\outrm{\mathrm{out}}
\newcommand\M{\mathbf{M}}
\newcommand\ro{\hat\rho}
\newcommand\half{\frac{1}{2}}

\begin{document}
\title{Non-Markovian Open Quantum Systems: Input-Output Fields, Memory, Monitoring}   
\author{Lajos Di\'osi}
\email{diosi@rmki.kfki.hu}
\homepage{www.rmki.kfki.hu/~diosi}
\affiliation{
Wigner Research Center for Physics\\
H-1525 Budapest 114, POB 49, Hungary}
\date{\today}

\begin{abstract}
Principles of monitoring non-Markovian open quantum systems are analyzed. 
We use the field representation of the environment
(Gardiner and Collet, 1985) for the separation of its memory and detector part, respectively. 
We claim the system-plus-memory compound becomes Markovian, the detector part is tractable 
by standard Markovian monitoring. Because of non-Markovianity, only the mixed state of the 
system can be predicted, the pure state of the system can be retrodicted. 
We present the corresponding non-Markovian stochastic Schr\"odinger equation.
\end{abstract}


\maketitle

In a seminal paper \cite{GarCol85} Gardiner and Collett used 
quantum white-noise and the related Markovian quantum field
to represent the dynamics of a quantum oscillator bath in the Markovian 
(memory-less) limit. This allowed the construction of exact stochastic  
differential equations to describe the influence of the bath B on the embedded (i.e.: open) 
quantum system S, the reaction of S on B, and the time-continuous
monitoring of S. The theory became standard 
in quantum optics \cite{GarZol04} and in many fields where a quantum system is open to natural or 
designed environmental influence \cite{WisMil10}. 
If the memory of B cannot be ignored for S then Markovian tools become jeopardized.
In non-Markovian (NM) case, S is coherently interacting 
with a finite part of B over a finite time.
From different theoretical efforts \cite{JacColWal99,JacCol00,GamWis03,Dio08a,WisGam08,Dio08b,Mazetal09} 
we distill a central question: how can we divide the environment B into 
the \emph{memory} M and \emph{detector} D? Part M is continuously entangled with S 
but the compound S+M becomes a Markovian open system, as we shall argue.
Part D contains information on S and can be continuously 
disentangled, i.e.: monitored, without changing the dynamics of S.

As a matter of fact, the Markovian field representation \cite{GarCol85} of 
B is capable to account for memory effects and leads 
to a natural separation between M and D. The local Markov 
field interacts with S in a finite range: this part makes 
the memory M. The output field carries away information on S, it makes the detector D.
Most features of the Markovian theory \cite{GarCol85} of monitoring apply invariably
to the composite system S+M. 

Earlier, Jack, Collett and Walls realized the role of a finite memory time in simulation \cite{JacColWal99}
and in monitoring \cite{JacCol00}. These authors calculated, for the first time, the retrodiction of the mixed
quantum state. Here we exploit the causality features of the standard Markovian bath \cite{GarCol85,GarZol04,WisMil10} 
and, among other results, calculate the retrodicted pure state and the current mixed state.    
 
Our work starts with a brief summary of the standard Markovian field theory \cite{GarCol85}. 
Then we identify the memory M and the detector D, and outline the scheme how  
S+M becomes a Markovian open system. Its Markovian master equation is derived followed 
by the derivation and discussion of the stochastic Schr\"odinger equation (SSE) of monitoring S. 

\emph{Markovian field, non-Markovian coupling.}
The composite S+B dynamics is based on the total Hamiltonian 
\begin{equation}\label{Htot}
\Ho=\Ho_S+\Ho_B+\Ho_{SB}, 
\end{equation}
where $\Ho_S$ is the Hamiltonian of S,
the bath Hamiltonian is $\Ho_B=\int\om\bo_\om\dg\bo_\om d\om$
and
$\Ho_{SB}=i\so\int\kappa_\om\bo_\om\dg d\om+\hc$
is their interaction,
where $\so$ is a S-operator that couples to the B-modes.
Here $\bo_\om$ are boson annihilation operators for the $\om$-frequency modes of B, 
satisfying $[\bo_\om,\bo_{\om'}\dg]=\delta(\om-\om')$.
B can be called Markovian because of the flat spectrum.  
Memory effects are fully encoded in the coupling $\kappa_\om$. 
If the coupling is frequency-independent, $\kappa_\om=\const$, then S is
Markovian open system, otherwise it has a memory. We are interested in the latter case,
i.e., in NM open systems. We assume that
S and B are initially uncorrelated. Let, for simplicity's, the initial B-state be
the vacuum $\vert0\rangle$ defined by $\bo_\om\vert0\rangle=0$ for all $\om$. 

We switch for an abstract field representation \cite{GarCol85,GarZol04,WisMil10}. 
The bath field $\bo(z)$ is defined by 
\begin{equation}\label{bz}
\bo(z)=\frac{1}{\sqrt{2\pi}}\int\bo_\om\e^{-i\om z}d\om,
\end{equation}
where $z$ is a real 1-dimensional spatial coordinate. For convenience, we set the velocity
of propagation to $1$. The canonical commutation relationship is local:
\begin{equation}\label{bzComm}
[\bo(z),\bo\dg(z')]=\delta(z-z'),
\end{equation}
hence the field can be measured \emph{independently} at all locations. In particular, it can
be measured in the coherent state overcomplete basis parametrized by the complex field $\xi(z)$.
The (unnormalized) Bargman coherent states
\begin{equation}\label{cohxidispl}
\vert\xi\rangle=\exp\left(\int\xi(z)\bo\dg(z)dz\right)\vert0\rangle
\end{equation}
form an overcomplete basis:
\begin{equation}\label{cohxicompl}
\M\vert\xi\rangle\langle\xist\vert=\hat1.
\end{equation}
$\M$ stands for the integral (mean) over $\xi$ with the normalized measure
according to the standard complex white-noise statistics, specified by
\begin{equation}\label{xi}
\M\xi(z)=0,~~\M\xi(z)\xi(z\pr)=0,~~\M\xi(z)\xi^\ast(z\pr)=\delta(z-z\pr).
\end{equation}   
If we perform the measurement, the state of B collapses on $\vert\xi\rangle$ randomly,
the complex field $\xi(z)$ becomes the random read-out. But its statistics depends on
the pre-measurement state. In the vacuum state $\vert0\rangle$, the read-outs
$\xi(z)$ follow the statistics (\ref{xi}). This statistics gets modified by 
the B-S interaction. Typically, the mean becomes non-vanishing, cf. (\ref{bouttcl}) or
(\ref{bouttclsel}).

Both the bath and interaction Hamiltonians can be written in terms of the fields:
\begin{equation}\label{H_Bz}
\Ho_B=\frac{i}{2}\int\bo\dg(z)\partial_z\bo(z)dz+\hc,
\end{equation}
\begin{equation}\label{H_SBz}
\Ho_{SB}=i\so\int\bo\dg(z)\kappa(z)dz+\hc,
\end{equation}
where $\kappa(z)$ is the Fourier transform of $\kappa_\om$.

The underlying picture \cite{GarCol85,GarZol04,WisMil10} is that all B-modes are spatial 
excitations along a single direction $z$. 
\begin{figure}
\includegraphics{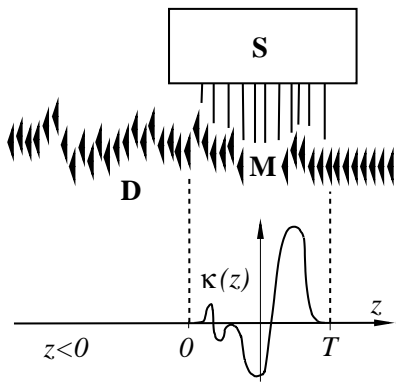}
\caption{The bath field $\bo(z,t)$, when free, is propagating from \emph{right to 
left} without dispersion at velocity $1$. The unperturbed
input field from range $z\geq T$ propagates through the interaction 
range $z\in[0,T]$ of non-zero coupling $\kappa(z)$,
gets modified by, and entangled with the system S, then it leaves to 
freely propagate away to \emph{left} infinity as the output field. 
The interaction range makes the memory $M$ and the output
range $z\leq0$ makes the detector D which can continuously be read out (monitored).} 
\label{nonmarkov_io_1}
\end{figure}
\begin{figure}
\includegraphics{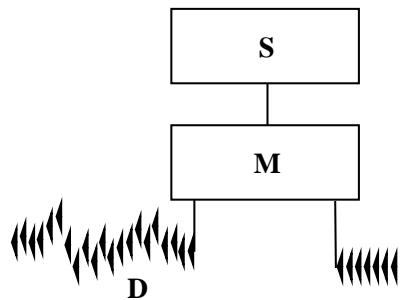}
\caption{If we form a memory subsystem M from the local field oscillators 
of the interaction range then the system S and the memory M constitutes
a Markovian open system. It is pumped by the standard Markovian quantum noise 
(input field) and it creates the Markovian output field D that can be monitored.}    
\label{nonmarkov_io_2}
\end{figure}
\begin{figure}
\includegraphics{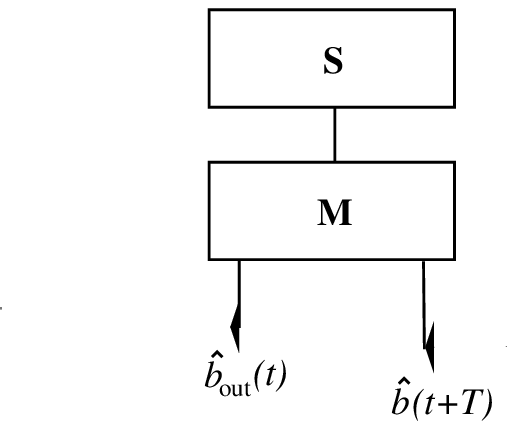}
\caption{The system-plus-memory is pumped by the standard (external) quantum
white-noise $\bo(t+T)$ and monitored through the modified quantum white-noise  
$\bo_\outrm(t)$ just like Markovian open quantum systems, apart form the delay
$T$ of read-out w.r.t. pump.}
\label{nonmarkov_io_3}
\end{figure}
The coupling $\kappa(z)$ is supposed
to vanish outside the \emph{interaction range}, say $z\in[0,T]$, where $T$ 
is the memory time. Memory effects are fully confined here. 
(If $\kappa(z)$ decays only asymptotically to zero, finite memory time can
still be a robust approximation \cite{JacColWal99,JacCol00}.)   
If $\kappa$ were a delta-function, $\kappa(z)\propto\delta(z)$, 
the interaction range would reduce to a single point $z=0$, 
memory effects would be absent and S would be Markovian open system.

\emph{Heisenberg picture.}
The solution of the Heisenberg field equation reads \cite{GarCol85,GarZol04,WisMil10}:
\begin{equation}\label{bzt}
\bo(z,t)=\bo(z+t)+\int_0^t\so(t-\tau)\kappa(z+\tau)d\tau.
\end{equation}
The first term $\bo(z+t)$ on the r.h.s. corresponds to free dispersionless propagation along the
line, from right to left (cf.~Fig.~\ref{nonmarkov_io_1}). The free field plays the role of the 
`conveyor belt' that carries information/perturbations one-way: from right to left,
never the opposite! As usual, the free field will later be identified as 
the field (\ref{bz}) in interaction picture:
\begin{equation}\label{bz_t}
\bo_t(z)=\bo(z+t).
\end{equation}
The second term on the r.h.s. of (\ref{bzt}) represents the interaction with S, 
localized inside the interaction range $z\in[0,T]$. 
In the \emph{input range} $z\geq T$ the vacuum field is freely propagating. 
In the \emph{output range}
$z\leq0$ the field is freely propagating and carrying \emph{away} the perturbations 
emerged in the interaction range. For $t>0$, the input field
does not depend on whatever happens at $z<T$, and the 
dynamics of S remains undisturbed whatever happens at $z\leq0$ to the output 
field. Most importantly, we can continuously observe the output field without altering
the dynamics of S. Accordingly, the memory M will consist of the local field
inside the interaction range and the detector D will consist of the 
output field. We emphasize that the coupling of M to the rest of B is 
Markovian hence S+M becomes Markovian open system (Fig.~\ref{nonmarkov_io_2}). 
The full armory of Markovian continuous measurement theories \cite{GarCol85,GarZol04,WisMil10}, 
including the Ito-formalism, could be deployed---with some peculiarities though.

As we said, the D part of the field is the output field $\bo(z\geq0,t)$. 
The earliest location of monitoring is $z=0$ and it is common to introduce the notation
$\bo_\outrm(t)=\bo(0,t)$ and it is common to call it the output field:
\begin{equation}\label{boutt}
\bo_\outrm(t)=\bo(t)+\int_0^t\so(t-\tau)\kappa(\tau)d\tau.
\end{equation}
This is the famous input-output relationship which works for the NM case as well.
The equation expresses the variable $\bo_\outrm(t)$ which
one can continuously monitor without affecting the dynamics of S. 
Since $\kappa(\tau)$ vanishes for $\tau<0$, the measured signal reflects delayed and
coarse-grained average of the S-variable $\so$. 
  
In particular, if we read out $\bo_\outrm(t)$ in ideal heterodyne measurement---which 
corresponds to the measurement in the coherent state basis (\ref{cohxidispl})---the resulting signal
$b_\outrm(t)$ contains the standard complex white-noise (\ref{xi}):
\begin{equation}\label{bouttcl}
b_\outrm(t)=\xi(t)+\int_0^t\langle\so(t-\tau)\rangle\kappa(\tau)d\tau,
\end{equation}
where $\langle\so(t)\rangle$ is the quantum expectation value of the Heisenberg operator. 

\emph{Markovian master equation.}
We construct the formal Markovian reduced dynamics of S+M in Schr\"odinger picture.
The Hamiltonian of M and the interaction are just $\Ho_B$ and
$\Ho_{SB}$, resp., restricted for the interaction range:
\begin{equation}\label{H_Mz}
\Ho_M=\frac{i}{2}\int_0^T\bo\dg(z)\partial_z\bo(z)dz+\hc,
\end{equation}
\begin{equation}\label{H_SMz}
\Ho_{SM}=i\so\int_0^T\bo\dg(z)\kappa(z)dz+\hc
\end{equation}
We are not ready yet. The outer input field $\bo(z>T)$ , that we cut off, will be replaced 
by the time-dependent vacuum white-noise $\bo(T+t)$ which is \emph{external} w.r.t. M since we take $t>0$.
This noise couples to $\bo(T)$ of the upper edge $z=T$ of M and pumps M via the following Hamiltonian: 
\begin{equation}\label{H_pump}
\Ho_{Mt}=i\bo\dg(T)\bo(T+t)+\hc
\end{equation}
(This choice can be confirmed in Heisenberg picture: the field equation 
$d\bo(z)/dt=i[\Ho_M+\Ho_{Mt},\bo(z)]$ yields
the correct solution (\ref{bz_t}) for $z\in[0,T]$.)
As to the output field, we trace out the modes for $z<0$
while we must retain $\bo(z=0)=\bo_\outrm$ if monitoring is included. 
The total Hamiltonian is $\Ho_S+\Ho_M+\Ho_{Mt}+\Ho_{SM}$. 
We can directly write down the corresponding master
equation for the density matrix of S+M:
\begin{eqnarray}\label{ro_SM}
\frac{\ro_{SM}}{dt}\!\!=\!\!&-&\!\!i[\Ho_S+\Ho_M+\Ho_{SM},\ro_{SM}]\\
                   \!\!&+&\!\!\bo(T)\ro_{SM}\bo\dg(T)-\half[\bo\dg(T)\bo(T)\ro_{SM}+\hc].\nonumber
\end{eqnarray}
The non-Hamiltonian term on the r.h.s. is the typical second-order contribution of 
the white-noise $\Ho_{Mt}$. 

We have thus transformed the original NM open system S into a standard
Markovian open system $S+M$ which is pumped by the vacuum 
white-noise $\bo(T+t)$ and could be monitored through $\bo_\outrm$, cf.~Fig.~\ref{nonmarkov_io_3}. 
In principle, the Markovian master equation (\ref{ro_SM}) would be a possible starting point to include
monitoring. Unfortunately, the obtained equation is formal, its application would require further
specifications on $\Ho_M$ regarding the boundary conditions. Rather we choose an alternative tool. 

\emph{Stochastic Schr\"odinger equation.}
We are interested in the dynamics of the monitored quantum state.
Non-Markovian SSEs \cite{DioStr97,DioGisStr98,StrDioGis99} are in tradition to describe open system dynamics,
whereas their role in monitoring either was ignored \cite{GasNag99,Bud00,Basetal,Rodetal09,VegAloGas05,JinYu10} 
or urged for investigations \cite{JacColWal99}, then it led to difficulties \cite{GamWis03,Dio08a,WisGam08,Dio08b}. 
The difficulties, related to the causal relationship between S and D, become transparent 
in our new treatment. 

We shall work in the interaction picture: according to (\ref{bz_t}) we replace
$\bo(z)$ by $\bo_t(z)=\bo(z+t)$ and we replace $\so$ by $\so_t$.
The interaction (\ref{H_SMz})
becomes the functional of the standard vacuum white-noise $\bo(t)$:
\begin{equation}\label{H_t} 
\Ho_t=i\so_t\int_0^T \bo\dg(t+\tau)\kappa(\tau)d\tau+\hc
\end{equation}
In interaction picture the separate pump Hamiltonian (\ref{H_pump}) is not needed.
To construct the Schr\"odinger dynamics of S+M, 
let $\vert\Psi_S(0)\rangle$ stand for
the initial state of S and $\vert0\rangle$ for the initial vacuum state of M.
We choose an uncorrelated composite initial state:
\begin{equation}\label{PsiSM0}
\vert\Psi_{SM}(0)\rangle=\vert\Psi_S(0)\rangle\vert0\rangle.
\end{equation}
Using (\ref{H_t}), we get the following Schr\"odinger equation:
\begin{equation}\label{Sch}
\frac{d\vert\Psi_{SM}(t)\rangle}{dt}
=\left(\so_t\int_0^T \bo\dg(t+\tau)\kappa(\tau)d\tau-\hc\right)\vert\Psi_{SM}(t)\rangle.
\end{equation} 
Observe that the r.h.s. depends on the field $\bo(t+\tau)$ for $\tau\in[0,T]$, 
i.e., for later times than $t$ itself.

Like in case of Markovian open systems, we have to match the unitary evolution (\ref{Sch}) with the
continuous read-out of $\bo(t)$. To this end, we project the M-part of the composite state 
$\vert\Psi_{SM}(t)\rangle$ on the coherent state basis $\{\vert\xi\rangle\}$, cf.(\ref{cohxidispl}):
\begin{equation}
\vert\Psi_{S}[\xist]\rangle=\langle\xist\vert\Psi_{SM}(t)\rangle.
\end{equation}
The Schr\"odinger equation (\ref{Sch}) reads:
\begin{eqnarray}\label{Schst}
\frac{d\vert\Psi_S[\xist;t]\rangle}{dt}&=&
\so_t\int_0^T\!\!\!\!\!\!d\tau\kappa(\tau)\xist(t+\tau)\vert\Psi_S[\xist;t]\rangle\nonumber\\
&-&\so_t\dg\int_0^T\!\!\!\!\!\!d\tau\kappa^\ast(\tau)\frac{\delta\vert\Psi_S[\xist;t]\rangle}{\delta\xist(t+\tau)}.
\end{eqnarray} 
This equation is just the Schr\"odinger equation (\ref{Sch}) in different representation.
But it is more than that if we consider the monitoring and read-out of $\bo(t)$. Then
$\vert\Psi_{S}[\xist]\rangle$ is the (unnormalized) conditional state vector of S, depending on the
measured signal $b_\outrm=\xi$. Since the signal is stochastic, we call (\ref{Schst}) 
the non-Markovian SSE. 

We have come to a landmark.
The r.h.s. would contain the measured signal $\xi(t+\tau)$ at later times w.r.t. $t$, these data are
not yet available at time $t$.
We can still exploit the SSE in two ways. Either we propagate the \emph{conditional mixed state},
or we propagate the \emph{retrodicted pure state}. 
In both cases, we prepare the initial state (\ref{PsiSM0}) of S+M at time $t=0$, let it go and
start to read out the signal $b_\outrm(t)=\xi(t)$. The field in
M becomes entangled with S so we can never monitor the pure state $\vert\Psi_S\rangle$ of S.    
Nonetheless, at each time $t>0$ we propagate (calculate) the solution of (\ref{Schst}) by using the latest
read-outs $\xi(t)=b_\outrm(t)$ and by setting auxiliary values for $\xi(t+\tau)$ for $\tau\in(0,T)$. 
These latter data are not yet measured, 
we acknowledge our ignorance by tracing out the corresponding field degrees of freedom. 
Accordingly, we derive the following conditional mixed state from the pure state solution:
\begin{equation}\label{roSSE}
\ro_S[b_\outrm,b_\outrm^\ast;t]=
\M\vert\Psi_{S}[\xist;t]\rangle\langle\Psi_{S}[\xi;t]\vert_{\xi(\sigma<t)\equiv b_\outrm(\sigma<t)}.
\end{equation}
This mixed state (with a normalizing factor) is the true conditional state of S under monitoring. 
If we stick to the idea of a conditional pure state, we exploit the measured signal $b_\outrm(t)$ differently. 
We use the SSE (\ref{Schst}) at time $t$ to retrodict the state propagation at time $t-T$.
Until time $t=T$, measured data are not sufficient to retrodict any pure state.
From time $t=T$ on, we start to propagate the initial state $\vert\Psi_S(0)\rangle$, using the
signal $b_\outrm=\xi$ measured until time $t$. At each time $t>T$, 
we have $\vert\Psi_{S}[b_\outrm^\ast;t-T]\rangle$ as the solution of the SSE. 
And this (with a normalizing factor) is our retrodicted 
conditional pure state for S. 
The pure state $\vert\Psi_{S}[b_\outrm^\ast;t-T]\rangle$ looks a mere mathematical construction 
though it will appear---as it were the true state---in the expression (\ref{bouttclsel}) of 
the measured output signal.  
 
So far we have not determined the statistics of the signal $b_\outrm$. The candidate expression
(\ref{bouttcl}) does not resolve the selective evolution of S under monitoring. 
This selection is only given by the SSE (\ref{Schst}) together with its interpretation (\ref{roSSE}).
As we said before, the signal $b_\outrm$ would be the standard complex white-noise (\ref{xi}) 
of zero mean had we switched off the interaction. 
With the interaction on, the typical change is that the mean of $b_\outrm$ will be non-vanishing.
Lessons from the Markovian special case and the non-selective NM form
(\ref{bouttcl}) would suggest the following expression:
\begin{equation}\label{bouttclsel}
b_\outrm(t)=\xi(t)+\int_0^t\langle\so_{t-\tau}\rangle_{t-\tau}\kappa(\tau)d\tau,
\end{equation}
where $\langle\so_{t-\tau}\rangle_{t-\tau}$ is the quantum expectation value of $\so_{t-\tau}$
in the conditional mixed state $\ro_S[b_\outrm,b_\outrm^\ast;t-\tau]$
or, alternatively, in the conditional pure state $\vert\Psi_{S}[b_\outrm^\ast;t-\tau]\rangle$. 
We show later the second choice is the right one.

\emph{Structured bath}.
Non-Markovian open systems are often derived from Markovian coupling $\kappa_\om=1$ to
a NM bath of non-flat spectral density $\alpha_\om\geq0$.
A prototype of NM SSE was obtained in 1997 \cite{DioStr97}: 
\begin{equation}\label{Schst97}
\frac{d\vert\Psi_S[\astar;t]\rangle}{dt}=
\so_t\astar_t\vert\Psi_S[\astar;t]\rangle
-\so_t\dg\!\!\int_0^t\!\!\!\!d\sigma\alpha(t-\sigma)\frac{\delta\vert\Psi_S[\astar;t]\rangle}{\delta\astar_\sigma}
\end{equation} 
Here $\astar_t$ must be a (Gaussian) complex colored noise of zero mean and of correlation
\begin{equation}\label{a_tCorr} 
\M a_t\astar_\sigma=\alpha(t-\sigma),
\end{equation}
where $\alpha(t)$ is the bath correlation function, i.e.: the Fourier transform of $\alpha_\om$.
The interpretation of this equation drew permanent attention. Gambetta and Wiseman showed \cite{GamWis03,WisGam08} 
that no monitoring process exists for $\vert\Psi_S[\astar;t]\rangle$ itself. If, however, the support of $\alpha(t)$ 
is finite (there is a finite memory time) then the SSEs like (\ref{Schst97}) can predict the mixed 
conditional state at $t$ and retrodict the pure conditional state at t minus the memory time 
\cite{Dio08a,Dio08b}. Now we are in a position to unfold the causality structure of the SSE 1997: we rewrite
it into the form of the SSE (\ref{Schst}).

The point is that the said NM bath with Markovian coupling can equivalently be substituted
by the Markovian B with the NM coupling satisfying $\vert\kappa_\om\vert^2=\alpha_\om$. 
Precisely, if we solve
\begin{equation}\label{alpha} 
\alpha(t)=\int\kappa(t+\tau)\kappa^\ast(\tau)d\tau
\end{equation} 
for $\kappa(t)$ at condition $\kappa(\tau)=0$ for $\tau<0$ \cite{Cho24} then 
we can express $a_t$ through the standard complex white-noise (\ref{xi}):
\begin{equation}\label{axi} 
a_t=\int\xi(t+\tau)\kappa^\ast(\tau)d\tau.
\end{equation} 
By inserting this into (\ref{Schst97}), the resulting equation
coincides with the NM SSE (\ref{Schst}). Therefore the discussion and resolution
of the causality issue of monitoring, cf. our preceeding paragraph, can be directly
adapted to the old form of the SSE \cite{foo}.

Let's verify the Girsanov transformation $\xi(t)\Rightarrow b_\outrm(t)$
underlying our heuristic expression (\ref{bouttclsel}) of the output signal.
We exploit the Girsanov transformation $a_t\Rightarrow\tilde a_t$ accomplished
by (16) in \cite{DioGisStr98}:
\begin{equation}\label{aout} 
\tilde a_t= a_t + \int_0^t \alpha(t-\tau)\langle\so_{t-\tau}\rangle_{t-\tau}d\tau
\end{equation} 
where $a_t,\tilde a_t$ are related to $\xi(t),b_\outrm(t)$, resp., by the convolution (\ref{axi}). 
Let's arrange all terms on one side, apply (\ref{axi}) and insert (\ref{alpha}),
yielding
\begin{equation}\label{xout} 
\int\!\!\!\Big( 
\!b_\outrm(t+\sigma)-\xi(t+\sigma)-\!\!\int_0^t \!\!\!\!\langle\so_{t-\tau}\rangle_{t-\tau}\kappa(\tau+\sigma)d\tau
        \!\Big)\kappa^\ast\!(\sigma)d\sigma\!=\!0.
\end{equation} 
The removal of the outer convolution, legitimated at least when $\kappa_\om$ is nowhere zero, 
yields our result (\ref{bouttclsel}).
From \cite{DioGisStr98} we know that $\langle\so_{t-\tau}\rangle_{t-\tau}$ must be taken
in the retrodicted pure state $\vert\Psi_{S}[b_\outrm^\ast;t-\tau]\rangle$.  
Since pure state retrodiction needs a minimum
time delay $T$, the theoretical prediction (\ref{bouttclsel}) of the output signal 
$b_\outrm(t)$ can only be calculated at time $t+T$, i.e., at current time $t$
the latest statistical retrodiction concerns $b_\outrm(t-T)$. This restriction is
a typical quantum-non-Markovian feature. (However, the delayed access to the output 
signal is a common relativistic feature for any monitoring where the finite speed of 
light matters.)

\emph{Summary}.
We applied the well-known Markovian field representation of the environmental bath at
non-Markovian coupling to the embedded open system. We argued that the field in
the vicinity of the system plays the role of memory responsible for the non-Markovianity,
far from this vicinity it remains Markovian and subject to standard Markovian
theory of monitoring. We unfolded the abstract bath into the memory and the detector part.
Our work should initiate further investigations along these principles. 

A formal master equation has been derived just to confirm Markovianity of the 
system-plus-memory compound. We have derived a stochastic Schr\"odinger equation of the 
monitored system and pointed out its role in predicting the conditional mixed state 
and in retrodicting the  conditional pure state---in accordance with recent discussions and 
anticipations about the fenomenological stochastic Schr\"odinger equation of Strunz and the
author. 

We are aware of two inevitable perspectives.
First, a certain asymptotic Markovianity of the system-plus-memory reduced dynamics is 
readily seen for the Szeg\"o class of couplings \cite{Chietal10,Maretal11}. 
Investigations of asymptotic Markovianity should be extended for our class of couplings 
together with considering double-sided chain representations for both input and output
regimes, respectively. (An independent method to treat the memory has appeared just recently 
\cite{YanMiaChe11}.) Second, Ito differential and integral calculus should be deployed 
to improve our tentative derivations. 

Support by the Hungarian Scientific Research Fund under Grant No. 75129,
by the Bilateral Hungarian-South African R\&D Collaboration Project, by EU COST Action MP100,
and extensive discussions with Thomas Konrad and Francesco Petruccione are gratefully acknowledged.


\begin{thebibliography}{99}
\bibitem{foo} The old SSE \cite{DioStr97} was in `wrong' variables $a_t$ for monitoring. 
With the new variables $\xi(t)$ and with the field representation we have spared both the 
extensive use of functionals and the sophisticated construction of the time-continuous detector
in \cite{Dio08a,Dio08b}.
\end{thebibliography}
\end{document}